# Digital Libraries: From Process Modelling to Grid-based Service Oriented Architecture

Zaheer Abbas Khan
CCCS, Faculty of CEMS,
University of the West of
England, Bristol, UK
Zaheer2.khan@uwe.ac.uk

Mohammed Odeh
CCCS, Faculty of CEMS,
University of the West of
England, Bristol, UK
Mohammed.Odeh@uwe.ac.uk

Richard McClatchey
CCCS, Faculty of CEMS,
University of the West of
England, Bristol, UK
Richard.McClatchey@uwe.ac.uk

**Abstract**

*Graphical Business Process Modelling Languages (BPML) like Role Activity Diagrams (RAD) provide ease and flexibility for modelling business behaviour. However, these languages show limited applicability in terms of enactment over distributed systems paradigms like Service Oriented Architecture (SOA) based grid computing. This paper investigates RAD modelling of a Scientific Publishing Process (SPP) for Digital Libraries (DL) and tries to determine the suitability of Pi-Calculus based formal approaches to enact SOA based grid computing. In order to achieve this purpose, the Pi-Calculus based formal transformation from a RAD model of SPP for DL draws attention towards a number of challenging issues including issues that require particular design considerations for appropriate enactment in a SOA based grid system.*

## 1. Introduction

Business organisations are not just limited to specific localities and artefacts but they can gradually span geographical boundaries and can produce ranges of new artefacts as business products. This gradual expansion of business organisations also influences the currently executing business processes. And, as a part of strategic business management, business organisations tend to adapt technological solutions for not only effective execution of their processes but also for competitive advantage over their competitors. This gradual expansion is effectively supported by BPML, where it not only helps to visualise the underlying process behaviour but also highlights improvement techniques. For example, RAD being graphical BPML is easy to understand and better supports the above mentioned improvements [4]. But this modelling perspective for expanding business processes becomes more challenging when modelled business processes need to adapt to distributed computing as their underlying implementation infrastructure. And, graphical BPML, like RAD, can limit the scope of such adaptation [3].

Digital Libraries (DL) [1,7] can be regarded as emerging potential business applications that can fit the context of the above mentioned problem for two reasons. Firstly, a DL embodies all the necessary elements that business applications have, for example business process support and management, in relation to business artefacts such as preserved articles, policies, etc. Secondly, to incorporate multi-disciplinary knowledge, DLs potentially seek solutions for their deployment over grid-based Service Oriented Architecture (SOA) infrastructure [6]. Furthermore, despite the benefits and vast amount of research in grid computing, the emergence of this technology into the business sector is as yet limited. This may be attributed to the relatively recent emergence of this technology and the need for extensive testing and its impact on the business processes in driving business applications. Thus, deploying existing business processes on a grid infrastructure may be considered as a step towards greater use of grid computing in business applications. Modelling of DL processes using RAD and their enactment in this grid environment embodies the above mentioned problem and seeks to bridge the existing gap between business process models and grid computing.

In order to bridge this gap, formal approaches based on Pi-Calculus [23] have been utilised to apply RAD modelling dynamisms into grid based executable systems. This is suitable since Pi-Calculus, not only supports process communication but also provides mobility support in terms of handling emerging communication channels and the integration of new processes (often referred to as mobile processes) [14]. A translation mechanism, with the assumption that one RAD role represents one Pi-Calculus process, can help to associate RAD process models to Pi-Calculus models. Thus to demonstrate this new approach we have translated the "Scientific Publishing Process" of DLs modelled in RAD into a proposed Pi-Calculus model. Figure 1 shows the overview of the problem space with a possible application which mediates between grid, DL and RAD modelling and formal methods like Pi-Calculus which has led to a suitable problem solution. The next few sections represent the grid emergence and its application requirements followed by a translation mechanism from the RAD model to the Pi-Calculus model which is shown through an example model.

Finally, a discussion is presented on the outcomes or the issues raised by performing this translation based on which RAD models can lead to SOA based grid system deployments.

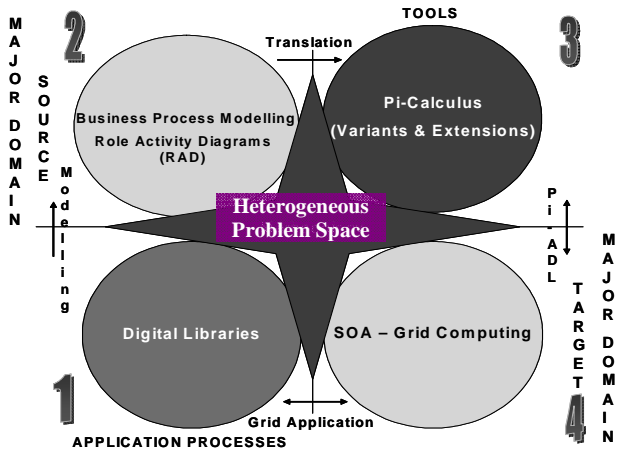

**Figure 1: Problem Space and Solution**

## 2. Grid Emergence and Process Modelling Requirement

The past few years have shown developments in grid technology which have not only introduced infrastructural changes (e.g. the recent shift from Open Grid Services Infrastructure (OGSI) to Web Services Resource Framework (WSRF)), but have also supported new application domains ranging from scientific experiments to e-Research, DL etc. Figure 2 shows these developments where the grid integrates several other technologies such as distributed or mobile Agents, Web services [8] etc., with underlying shared semantics, to support new application domains such as e-Research, e-Teaching etc. This continuous emergence potentially requires alternatives (as shown in Figure 3) for modelling the served systems processes [9] in a manner which can both reflect the changes from served system models to grid systems and/or can adapt emerging grid systems for served system process models. In this context, considering the DL process modelling, as a case study can help to understand the behaviour of DL processes and provide support to realise its enactment requirements over a SOA based grid system.

In order to support such an enactment, there can be more than one possibility to move from business process modelling towards SOA based grid systems. Figure 3 shows an abstract differentiation of two separate approaches which can be adapted for further study and can lay the foundation for experiments and evaluation. The first alternative is either to model or enhance the SOA based grid system to support or adapt the business processes. This might require the grid system modelling keeping inline with the business processes. The second alternative is to, extend the business process modelling languages to model business processes which can be enacted into a SOA based grid systems either directly or indirectly.

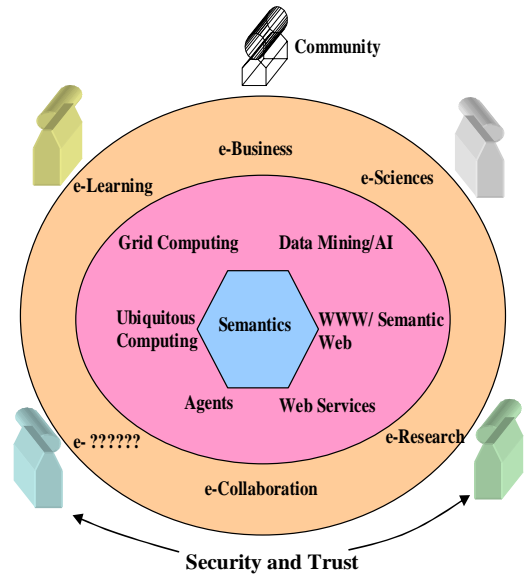

**Figure 2: Evolution and Future Perspective of Grid Computing**

The former approach requires extensions to system modelling languages, as proposed by the Rational Unified Process (RUP) and UML proponents mentioned in [2, 5], but it is less appealing since it might require infrastructural enhancements in grid systems, which will be more specific to a particular business process. Also, any change in business processes may require business modification or may raise additional service requirements in underlying grid systems. The latter approach provides an additional opportunity for business environments to evaluate their executable business processes without any refactoring of the grid systems. Furthermore, any changes into the original business processes can also be reflected into the stable grid systems.

Both approaches may follow conventional methods (e.g., limited decomposable mapping from FBPML to OWL-S [3] and/or RAD to UML activity model [16]) or formal methods (e.g., 'Spanish Fish Market' [24] and other work [22, 26], more specific to web services [10, 11]) for their execution. The benefits of formal methods are that the models can be verified and validated against the requirements mentioned in the source Business Process Models (BPM) and can generate automated code. The next section provides a more comprehensive overview of formal approaches for BPM. This latter approach may also be divided into direct or indirect translation for executable SOA based grid applications. Both approaches have pros and cons. Direct translation will lead directly into enacted grid application services as compared to the indirect approach where first BPM will be translated into a system model and then that system model can be used to generate executable code for grid application services. The direct approach requires semantic support in addition to a design control process either through OWL-S or formal languages to generate code for grid services. Whereas the indirect approach supports reusability of the generated system models and can also be used for many other purposes or application enactment in addition to just SOA based grid application services.

The latter approach justifies the answer to the question why do businesses need to model their processes based on SOA-grid models from scratch? It could be beneficial for businesses to analyse and evaluate their systems deployed on grid technology in order to obtain maximum benefit of low cost and scalable computing infrastructures.

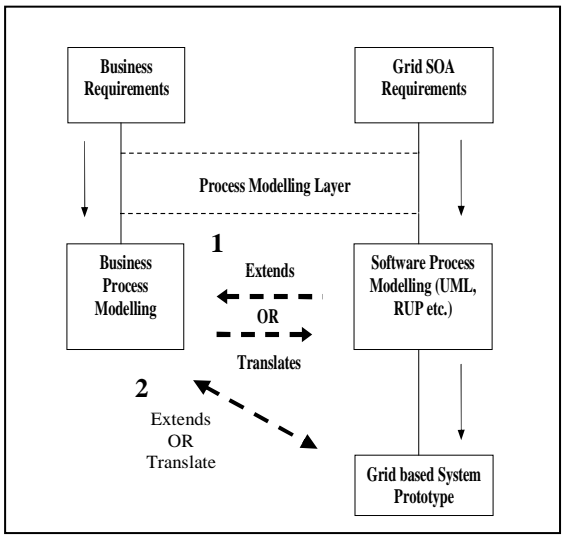

**Figure 3: Alternate Methodology**

## 3. Business Process Modelling Formalism

Normally, mathematical formalisms (e.g. Pi-Calculus) for BPML (e.g. RAD) can provide additional support to analyse, verify and validate business processes by using analysis tools such as RDT models [26]. We selected Pi-Calculus because it supports communication and a dynamic link configuration which might be needed for the development of dynamic architectures or to support the notion of process mobility [14] at later stages of the transformational development process. There are two schools of thought in this, where the first is supportive [14, 15, 19, 24, 26] and the other is against [13, 17, 18] the applicability of Pi-Calculus for business process modelling. Proponents of Pi-Calculus, such as Business Process Management, support the argument that the mobility property of Pi-Calculus suits the emergence of business processes. However its opponents criticise the tailoring of Pi-Calculus as the only viable solution and argue that the notion of Pi-Calculus is exaggerated and it might prove to be difficult to generalise Pi-Calculus based solutions to all business environments. This seems possible, but this argument deserves much more research especially by conducting practical experiments to evaluate the viability of Pi-Calculus for general business environments rather than taking a specific and limited workflow example of deployment in a specific organisation. In contrast, pi-ADL (Architecture Description Language, ADL), which is also based on Pi-Calculus, seems to be more attractive and provides additional tools to verify and validate formal abstract specifications representing the business processes and can generate Pi-Calculus code [25].

In this regard, research groups have been working in mapping different graphical business modelling techniques to process algebra [22], Pi-Calculus [19, 21, 26] and eXtensible Markup Language (XML) [20] etc. Most of this mapping is reflected from state-based models [22, 24, 26]. It is obvious that under certain assumptions the mapping between graphical modelling techniques and Pi-Calculus is possible. The only issue that needs to be addressed is the generality of the applied mapping rules to cover the vast range of graphical modelling notations. In addition to addressing this, our work which is an extension in this regard, also investigates the applicability of business processes over SOA based grid systems by evaluating the applicability of pi-ADL based solutions.

## 4. Mapping of Scientific Publishing Process RAD model to Pi-Calculus

We have adopted the second approach mentioned in section 2 and have modelled the Scientific Publishing Process (SPP) of DL through RAD diagramming, followed by a Pi-Calculus translation.

### 4.1. Scientific Publishing Process (SPP): RAD Model

Publishing[1] is a very complex and rich process which incorporates many other relevant sub-processes such as peer-review, editing etc., to improve the quality and to establish the relationship with DL. That is why it is difficult to limit the scope of SPP. The overall purpose was to understand the SPP in its context with DL and to discover the complexities involved in RAD modelling. As a first indication, figure 4 shows a limited view of the RAD model of SPP which mainly covers Researcher and Co-researcher roles and their interactions with other roles participating in SPP. In the next sub-section Figure 5 also provides a limited translation of Figure 4 into Pi-Calculus. The complexity and richness of SPP can be measured by considering the rest of the twelve other roles (simply shown in Figure 4 as empty role blocks) which participate in SPP with different responsibilities.

### 4.2. Scientific Publishing Process: Pi-Calculus Model

For a complete understanding of the Pi-Calculus model, users must go through the table 1 and additional pointers.

---

[1] The RAD model for SPP is in the process of submission to another conference.

This also identifies some limitations which still exist in RAD diagramming and its mapping into Pi-Calculus.
1. We tried to keep naming conventions simple and easy to understand. Each symbol/notation written in Pi-model implicitly reflects its appropriate roles, activity or interaction in the RAD models. However improved naming rules for roles, activities, states etc. are still required.
2. Each role of the RAD process model is mapped as a Pi-Calculus process
3. RAD supports both synchronous as well as asynchronous interactions among roles, but we still need to differentiate such interactions in the Pi model.
4. The a\b port identifies the participating roles in an interaction and the first identifier shows the initiating side in an interaction.
5. The a\b* interaction port represents a 2-way interaction initiated by a.
6. Multiple Roles involved in a single interaction can be identified by the a\b\c port.
7. Double lines over an activity represent encapsulation.
8. An activity followed by () represents an external event.
9. Each State/goal starts with upper case..
10. Each Activity starts with lower case,.
11. This Pi model does not show the cardinality and instantiation of specific roles.

Due to limited space availability, we show in Figure 5 just the Researcher and Colleague-Researcher roles modelled in Pi-Calculus.

**Table 1: RAD role to Pi-Calculus notations**

| No. | Scientific Publishing Process RAD Model Roles | Pi Calculus Symbols/ Notations |
|---|---|---|
| 1 | Researcher | Resr |
| 2 | Colleague/Co-Researcher | Colres |
| 3 | Publication Practice | Pubprac |
| 4 | Editor | Editr |
| 5 | Editorial Board | Editbrd |
| 6 | Multimedia Content Management | Mmcntmgt |
| 7 | Referee | Refr |
| 8 | Publisher | Publ |
| 9 | Repository | Repo |
| 10 | Backup Service | Backup |
| 11 | Metadata Cataloguer | Metacata |
| 12 | Bibliographic/Indexing Service | Bibl |
| 13 | Practitioner | Prac |
| 14 | Catalogue Repository | Catarepo |

## 5. Translation Limitations and Discussion

Pi-Calculus suits the RAD model translation into grid based systems for two principal reasons. Firstly, Pi-Calculus supports process communication and provides mobility support in terms of handling emerging communication channels as well as integrating new processes [23]. Furthermore, this is very similar and suitable for RAD modelling which involves many interactions among participating roles in a process. New interactions and roles are well supported by the Pi-Calculus mobility and process notions. Secondly, the emerging nature of underlying system architectures for grid computing poses challenges for the executable business processes which are running as services on top of the grid. Here, the use of formal languages in defining the system architectures and then business processes, can make them general, validated and verified by automated tools. Furthermore, it can also encapsulate the underlying system emergence from application processes due to the standard formal descriptions.

The translation mechanism, with the basic assumption that one RAD role represents one Pi-Calculus process, helps to translate RAD process models into Pi-Calculus models. Accordingly, and to demonstrate this new approach we translated the SPP of DLs modelled in RAD to its proposed Pi-Calculus model, as shown in Figures 4 and 5. This translation can be easily seen as straightforward with few exceptions such as dealing with external events, state merges, loops, multiple role instances, etc., which might require additional or suitable representation mechanisms in the Pi-Calculus model. However, this straightforward translation is still not associated with an executable model which can be deployed in grid based SOA environments. However, this limitation can be recovered using pi-ADL as its architectural descriptions can be executed directly with the most significant add-on of generating a formalised grid-based hosting environment enabled by specifications for core grid services.

It is obvious that any formal model based on process algebra faces challenges when dealing with Service Oriented Computing (SOC) peculiarities like loose coupling, communication latency and open-endedness [12]. In addition, it has become evident after having analysed business processes modelled in RAD that the above simple translation from RAD process models to corresponding Pi-Calculus cannot be followed without shortcomings due to two main challenges. Considering the first of these challenges,, modelling in RAD is quite flexible and adopts the role-based approach where concrete models normally involve human interactions and encapsulate certain atomic activities to make these models simpler. On the one hand, this modelling approach significantly simplifies the complexities involved in business transactions. On the other hand, this flexibility poses some restrictions on finding mechanisms to relate RAD roles to grid services. A RAD role may contain several fine grain activities, which will require support from the current (or new) grid core services e.g., security, scheduling, information monitoring, and access services protocols like SOAP, GridFTP etc. Furthermore, these activities require separate grouping into different business application services running over those grid core services to maintain the separation of concerns and minimize principles. RAD models also require a certain level of granularity, either at the functional level or the activity level in RAD roles, in order to properly support a translation mechanism. Furthermore, this requires identification and elimination of manual activities that exist in RAD business process models which cannot be supported by computerised systems. It is also not wise to overindulge state or goal notation in RAD process models, as it increases the complexity in the models. Although, it is not RAD modelling limitation but the inherent tendency of synchronising the translations from graphical models to formal models [22, 24], make excessive use of this notation. In conclusion, this has inspired current research to work on the possibility of extending the RAD process modelling notation to support the later enactment in grid-based SOA environments. This will facilitate the smooth translation from RAD models to grid-based systems as a result of automating the generation of executable code.

The second challenge in translating RAD process models to corresponding Pi-Calculus can be mainly attributed to the evolving hybrid infrastructure of the grid, which necessitates the separation of business process models from their grid-based environment. Whereas the inherent service orientation tends to wrap such hybrid behaviour, making it easier and adaptable for business process execution, it is difficult to comment at this stage about the implication of such hybrid behaviour and the constructive or destructive process interactions with the underlying core grid services.

The simple translation from RAD process models to corresponding Pi-Calculus models will significantly benefit from properly addressing these two challenges.

## 6. Conclusions

Graphical BPMLs like RAD are effective in understanding, analysing and improving business processes. They can help in deploying business processes over SOA based grid systems. However the RAD modelling of SPP for DL shows that such modelling is limited in terms of its applicability over SOA based grid computing. Also, despite the benefits of using a formal language, which is based on Pi-Calculus, additional design considerations are needed to mask its simplicity and satisfy the complexity of SOA based grid systems. In this regard, we showed that it is possible to translate RAD process models into Pi-Calculus. This translation is however not without shortcomings and extensions in RAD notations are needed with certain design decisions to enable the mapping between RAD and Pi-Calculus. In addition, future work will be more focused on the applicability of pi-ADL in bridging the gap between business process models based on RAD diagramming and SOA based grid systems.

coupling as good design

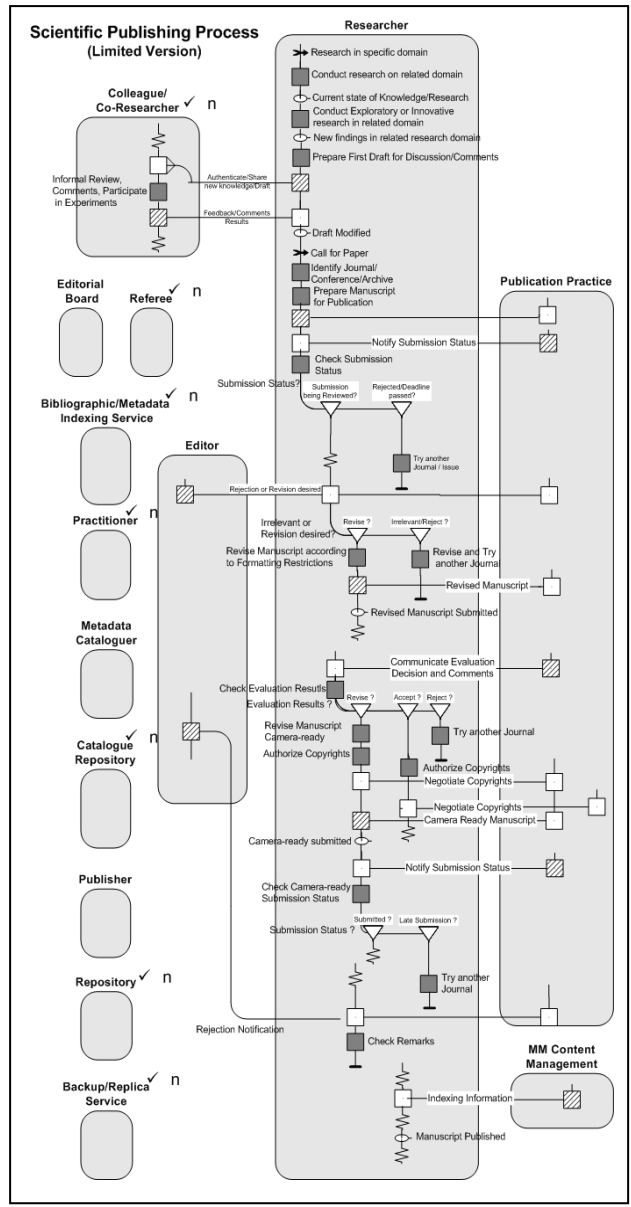

## 7. References

[1] Gio Wiederhold, "Digital Libraries, Value, and Productivity", Communications of ACM, April 1995/Vol. 38, No. 4, p.p. 85-96.

[2] Odeh M., et. al., "Bridging the Gap between Business Models and System Models", Information and Software Technology, Special Edition on Modelling Organisational Processes, Vol. 45, Issue 15, pp 1053-1060, Dec. 2003.

[3] Guo L. et. al. "Mapping a Business Process Model to a Web Services Model". Proceedings of the Third International Conference on Web Services, ICWS 2004, San Diego, California, USA, July 6th-9th, 2004.

[4] Ould M., "Business Processes: Modelling and Analysis for Re-engineering and Improvement", ISBN # 0-471-95352-0, 1995.

```
A. Scientific Publishing Process: Pi-Calculus representation:
SPP (resr\colres*, resr\pubprac*, edit/resr, mmcmtmgt/resr, . resr\pubprac*,
pubprac\edit*, pubprac\mmcrtmgt, pubprac\pub, pubprac\editbrd1,... edit\pubprac*,
edit\editbrd* , edit\resr, ref\edit )

SPP = def  Resr || Colres || Pubprac || Editr | Editbrd | Mmcmtmgt || Refr || Publ || Repo
 || Metacata || Bibl || Prac || Catarepo || Backup

A.1. Researcher:
Resr(resr\collres*, resr\pubprac*, edit/resr, mmcmtmgt/resr) = def risd().crord.Csok
Csok = cerrd.Nfird
Nfird = pfdfd.resr\collres < draft > col\res\resr (mdraft, comments).Dm
Dm = cp().ij.ca.pmfb. resr\pubprac < manuscript >.pubprac\resr (status).css
    ⎧ if ( submission = late)
    |     taj.stop()
    | else if (submission = reviewed)
    ⎩     time().edit\resr (rstatus)

    ⎧ if (rstatus = reject)
    |     ralaj.stop()
    | else if (rstatus = revise)
    ⎩     rmaff.resr\pubprac < manuscript >. Rms.time()

Rms = pubprac\resr (evaluation, comments) cer
    ⎧ if ( evaluation = reject)
    |     taj.stop()
    | else if (evaluation = accept)
    |     acr.resr\pubprac* (copyright).time()
    | else if (evaluation = revise)
    ⎩     rmcr.acr.resr\pubprac* (copyright).resr\pubprac < camera-ready >.Crsub

Crsub = pubprac\resr(sstatus).ccrss
    ⎧ if ( sstatus = late)
    |     taj.stop()
    | else if (sstatus = submit)
    ⎩     time()

    .{time().mmcmtmgt\resr(index).time().Mp.time()} + {edit/resr/pubprac(notify) cr.stop()}

Mp = time().done()

A.2. Colleague/Co-Researcher:
Colres ( colres\resr *) = def  time().resr\colres(draft).ircpie.colres\resr <mdraft,
comments >.time()
```

**Figure 5: Pi-Calculus translation for Researcher and Co-Researcher Roles in Scientific Publishing Process**

[5] Geoffrey Sparks, "The Business Process Model: An Introduction to UML", An introduction to modelling software systems using the Unified Modelling Language: The Business Process Model, Enterprise Architect UML Case Tool by Sparx Systems, 2000. URL access: http://www.sparxsystems.com.au/

[6] Frommholz I. et. al., "Supporting Information Access in Next Generation Digital Library Architectures", Digital Library Architectures: Peer-to-Peer, Grid, and Service-Orientation. Proceedings of the Sixth Thematic Workshop of the EU Network of Excellence DELOS, p.p. 49 – 60, 2004.

[7] Marcos A. G., Edward A. Fox, Layne T. watson, Neill A. Kipp, "Streams, Structures, Spaces, Scenarios, Societies (5S): A Formal Model for Digital Libraries" ACM Transactions on Information Systems, Vol. 22, No. 2, April 2004, p.p. 270 – 312.

[8] Khan Z. et. al., "A Semantic Grid based E-learning Framework (SELF)", ISBN: 0-7803-9075-X, Collaborative and Learning Applications on Grid Technology (CLAG), 5th International Symposium on Cluster Computing and Grid Cardiff, Wales, UK, May 2005.

[9] MC Winter, DH Brown, PB Checkland, "A role for soft systems methodology in information systems development", European Journal of Information Systems, 1995

[10] Mariya Koshkina and Franck van Breugel, "Verification of business processes for Web Services", Technical Report: CS-2003-11, York University, Canada. 2003.

[11] Andrea Ferrara, "Web Services: a Process Algebra Approach", 2nd International Conference on Service Oriented Computing, ICSOC04, New York City, NY, USA, November 15-18, 2004.

[12] M. Bravetti, G.Zavattaro, "Service Oriented Computing: a new challenge for Process Algebras", Short Contributions from the Workshop on 'Algebraic Process Calculi: The First Twenty Five Years and Beyond', PA '05, Bertinoro, Forl`ı, Italy, August 1–5, 2005, BRICS Notes Series.. NS-05-3, ISSN 0909-3206, June 2005.

[13] Wil M. P. van der Aalst, "Why workflow is NOT just a Pi-process", BPTrends, February 2004.

[14] Sanin Saracevic, "Business Process Management: Process-Oriented Paradigm", Report: CIS 8130 Spring 2004.

[15] Howard Smith, Peter Fingar, "Workflow is just a pi process", accessible at Site URL:

http://www.bpmi.org/bpmi-library/2B6EA45491.workflow-is-just-a-pi-process.pdf.

[16] M. Odeh, I. Beeson, S. Green and J. Sa, "Modelling Processes Using RAD and UML Activity Diagrams: an Exploratory Study", The 3rd International Arab Conference on

Information Technology, ACIT2002, December 16-19, Doha.

[17] Jon Pyke and Roger Whitehead, "Does Better Math Lead to Better Business Processes?", November 2003, BPTrends 2003

[18] W.M.P. van der Aalst, "Pi calculus versus Petri nets: Let us eat "humble pie", rather than further inflate the "Pi hype", BP Trends, 2005.

[19] Yang Dong, Zhang Shen-sheng, "Approach for workflow modelling using pi-Calculus", ISSN 1009 – 3095, Journal of Zhejiang University SCIENCE V. 4, No. 6, pp 643 – 650, Nov. – Dec. 2003.

[20] Martinez, A.I. and Mendez, R., "Integrating Process Modeling and Simulation Through Reusable Models in XML", In Proceedings of the Summer Computer Simulation Conference 2002, The Society for Modeling and Simulation International-SCS, ISBN 1-56555-255-5. pp. 452-460, Julio del 2002.

[21] Katerina Korenblat and Corrado Priami, "extraction of pi-calculus specifications from uml sequence and state diagrams", Technical Report 2003, University of Toronto.

[22] Costin Badica, Chris Fox, Business Process Modelling Using Process Algebras, presented at OR'43 (BPCME'01), 2001, Bath, UK

[23] Joachim Parrow, "An Introduction to the pi-Calculus", In Handbook of Process Algebra, ed. Bergstra, Ponse, Smolka, pages 479-543, Elsevier (2001).

[24] Julian A. Padget, Russell J. Bradford, "A pi-calculus Model anish Fish Market", Preliminary Report.

AMET 1998: p.p. 166-188.

[25] Flavio Oquendo, "pi-ADL: An Architecture Description Language based on the Higher-Order Typed pi-Calculus for specifying Dynamic and Mobile Software Architectures", ACM Software Engineering Notes, Volume 29, Number 4, May 2004.

[26] Robert John Walters, "Checking of models built using a graphically based formal modelling language", The Journal of Systems and Software 76 (2005) p.p. 55-64.